%% file: TWt.tex
\pdfoutput=1

\documentclass[aps,prl,twocolumn,superscriptaddress,groupedaddress]{revtex4} 

\usepackage{graphicx}  
\usepackage{dcolumn}   
\usepackage{yfonts}
\usepackage{bm}        
\usepackage{amssymb}   
\usepackage{amsmath}
\usepackage{slashed}
\usepackage{ytableau}
\usepackage{hyperref}
\usepackage[capitalize]{cleveref}

\hypersetup{colorlinks=true,linkcolor=magenta,anchorcolor=green,citecolor=blue,filecolor=black,menucolor=black,urlcolor=blue}

\usepackage{cleveref}
\begin{document}
\widetext

\title{Towards Gravity From a Color Symmetry}
\input author_list.tex      
\begin{abstract}
Using tools from color-kinematics duality
we propose a holographic construction of gravitational amplitudes, based on a
2d Kac-Moody theory on the celestial sphere. In the $N\to \infty$ limit the gauge group corresponds to $w_{1+\infty}$, due to the $U(N)$ generators enjoying a simple quantum group structure, which is in turn inherited from a twistor fiber over the celestial sphere. 

We show how four-dimensional momentum-space is emergent in this picture, which connects directly to the so-called kinematic algebra of the tree-level S-Matrix. On the other hand, the framework can be embedded within a celestial CFT to make contact with holographic symmetry algebras previously observed in the soft expansion. Kac-Moody currents play the role of a graviton to all orders in such expansion, and also lead to a natural notion of Goldstone
modes for $w_{1+\infty}$. Focusing on MHV amplitudes, main examples are a BCFW type recursion relation and holomorphic three-point amplitudes.
\end{abstract}
\maketitle

\section{Introduction}

Diverse aspect of gravitational theories suggest that
their dynamics enjoy integrable properties. For one, the relation between (super)gravity and (super)YM theories
via the \textit{double copy/color-kinematics duality} \cite{Bern:2019prr} suggests certain that avatars of Yangian symmetry \cite{Dolan:2004ps,Drummond:2009fd,Arkani-Hamed:2012zlh}
can emerge in perturbative gravity amplitudes \cite{Bjerrum-Bohr:2008qoa,Bern:2005bb,Bern:2018jmv}. Classical solutions of interest, such as the
Kerr black hole and its deformations, also enjoy hidden symmetries reflecting integrability in their dynamics \cite{Frolov:2017kze,Buonanno:1998gg}. The relation between such classical and quantum notions of integrability is a topic of active research unearthing intriguing connections, see e.g. \cite{Caron-Huot:2014gia,Caron-Huot:2018ape,Aminov:2020yma,Bianchi:2021mft}.

On a parallel venue, the celestial holography program \cite{deBoer:2003vf,Cheung:2016iub,Pasterski:2017kqt,Pasterski:2016qvg,Pasterski:2021rjz,Fan:2022vbz} aims to reformulate scattering in asymptotically flat spacetimes as governed by a putative CFT in the celestial sphere. The defining recipe is to recast the four-dimensional S-Matrix in a boost eigenstate basis where it can
be interpreted as a correlation function. From a pragmatical perspective it is expected that celestial correlation functions can be determined with help of the constraints from conformal symmetry and extensions thereof, alleviating the need for quantum gravity computations. A first motivation for this was the realization of universal soft theorems in gravity in the language of the celestial CFT (CCFT) \cite{Donnay:2018neh,Puhm:2019zbl,Guevara:2019ypd,Adamo:2019ipt}. By pushing the soft analysis to arbitrary orders within the conformal framework \cite{Guevara:2019ypd}, 
recently a novel
hierarchy based on the $w_{1+\infty}$ symmetry algebra has been undercovered in the
gravitational CCFT \cite{Guevara:2021abz,Strominger:2021lvk}. This algebra and its quantization have a long history with deep connections to integrability \cite{https://doi.org/10.48550/arxiv.1211.1287,https://doi.org/10.48550/arxiv.1305.4100,Prochazka:2015deb,hoppe2008lectures} and Penrose's twistor construction \cite{Penrose:1976jq,Penrose:1987uia}. Because of this, it is pressing to understand its implications in the gravitational S-Matrix, directly in momentum space, investigating to what extent scattering is constrained. 

In this letter we take a first step in this direction by introducing a 2d Kac-Moody theory that reproduces certain gravitational amplitudes through a celestial dictionary. Following the lore of color-kinematics duality, we show how four-dimensional momentum space emerges as the $w_{1+\infty}\cong U(N\to \infty)$ color group of the theory, thus explicitly manifests the hierarchy in the S-Matrix language.

\section{From the S-Matrix to $w_{1+\infty}$}

To present the idea we first point out that the algebra has a universal
imprint in the S-Matrix via collinear factorizations. In perturbative gravity amplitudes, such are given by splitting functions which also persist in a class of loop  amplitudes \cite{Bern:1998xc,Ball:2021tmb}, nevertheless here we will focus on the tree-level instance.

Consider the scattering
amplitude $\mathcal{A}_n(p_i)$ dressed with momentum conservation. For massless particles
collinear singularities correspond to the three-point on-shell vertices
which are unique for given helicities \cite{Benincasa:2007xk}. We consider only helicity-preserving vertices
whose coupling $16\pi G=1$ is further fixed by the equivalence principle, see Appendix II for the generic case. The collinear limit is then universal and reads
\begin{align}
\mathcal{A}_{n{+}1}(k^{+2}&,p_{1}^{\pm h},\ldots,p_{n})\nonumber\\
& \rightarrow\frac{(\epsilon^{+}\cdot p_{1})^{2}}{k\cdot p_{1}}\mathcal{A}_{n}((k{+}p_{1})^{\pm h},\ldots,p_{n}) \,,\label{eq:agsd}
\end{align}
We have emphasized the helicities through superscripts $p^{\pm h}$.
Hereafter we use $(2,2)$ signature where the Lorentz group is $SL(2,\mathbb{R})_z\times SL(2,\mathbb{R})_{\bar{z}}$ \cite{Atanasov:2021celesttorus}: We write momentum vectors in
terms of spinors as \cite{Elvang:2013cua}
\begin{equation}
p_{i}=|\eta_{i}\rangle[\lambda_{i}|\,\,\,\,,  k=|\eta\rangle[\lambda| \,\,\,,
\end{equation}
so that
$|\eta_{i}\rangle=(1,z_{i}),$ and $|\lambda_{i}]$ is an independent
spinor carrying the energy scale. Similarly for the graviton momenta $k$, which has coordinate e.g. $z_p$ and $|\lambda]$. Our collinear limit is then $z_p\to z_{1}$
where $\lambda,\lambda_{1}$ are kept generic. This yields $k+p_1 = |\eta_1\rangle ([\lambda|+[\lambda_1|)$. With a concording
change of notation, eq. \eqref{eq:agsd} can be written
\begin{equation}
\mathcal{A}_{n}^{\lambda\lambda_{1}\cdots\lambda_{n}}(z_{p},z_{1},\ldots,z_{n})\sim \frac{[\lambda \lambda_1]}{z_{p1}}\mathcal{A}_{n}^{\lambda_{1}+\lambda,\cdots\lambda_{n}}(z_{1},\ldots,z_{n})\,, \label{eq:gw}
\end{equation}
with $[\lambda \xi]=\epsilon^{\alpha\beta}\lambda_{\alpha}\xi_{\beta}$. In the following we will reinterpret this as a $SL(2,\mathbb{R})_z$
covariant OPE associated to a graviton operator and a massless helicity-$h$ field:
\begin{equation}
\mathbf{G}^{\lambda} (z_{p})\mathbf{O}_{h}^{\lambda_{1}}(z_{1})\sim\frac{[\lambda \lambda_1]}{z_{p1}}\mathbf{O}_{h}^{\lambda+\lambda_{1}}(z_{1})\label{eq:mfsad}
\end{equation}
where we emphasize that the antichiral components $\lambda,\lambda_{1}$
carrying `$SL(2,\mathbb{R})_{\bar{z}}$' weight are not colinear but generic.
Because of this the OPE holds not only in the soft sector but for
generic energies. In turn, it will become clear that writing the OPE
(\ref{eq:mfsad}) allows us to identify $\mathbf{G}^{\lambda_{p}}(z_{p})$
as Kac-Moody currents for a $w_{1+\infty}$ gauge group, where the
adjoint representation is paramatrized by $|\lambda]$, i.e. the kinematic
data! We will see this provides a precise realization of a 2d color-kinematics duality à la \cite{Cheung:2022vnd}. Furthermore the OPE suggests that scattering amplitudes
such as (\ref{eq:gw}) are then computed as correlations of such currents,
as we will discuss.

\section{$w_{1+\infty}$ as a color group}

Historically, the $w_{1+\infty}$ algebra and its quantum version
$\mathcal{W}_{1+\infty}$ arised as vertex algebras of higher-spin
fields in a 2d CFT, see \cite{Pope:1991ig,Prochazka:2015deb} for a review. In turn, the novel OPE (\ref{eq:mfsad}) is written
simply in terms of spin-1 currents and so obscures this interpretation.
It turns out that the connection can be made through the language
of quantum groups. To see this let us first provide a canonical definition
of $w_{1+\infty}$. Consider two abstract generators $\mu_{\pm}$
satisfying
\begin{equation}
[\mu_{\alpha},\mu_{\beta}]=i\tau\epsilon_{\alpha\beta}\,\,\,,\,\epsilon_{+-}=1\,,\label{eq:gsa}
\end{equation}
where $\tau$ is a quantization parameter such that as $\tau\to0$
this reduces to a Poisson bracket. The generators posses an obvious
$SL(2,\mathbb{R})$ symmetry $\mu_{\alpha}'=\Lambda_{\alpha}^{\beta}\mu_{\beta}$.
In this language the $w_{1+\infty}$ algebra emerges from the \textit{universal
enveloping algebra} of (\ref{eq:gsa}) (sometimes called Weyl algebra),
i.e. that given by totally symmetric polynomials of the operators
\begin{equation}
W^{s}_{\alpha_{1}\cdots\alpha_{s}}:=\mu_{(\alpha_{1}}\cdots\mu_{\alpha_{s})}\,\,,\,\,\alpha_{i}=\pm \,,\label{eq:2dws}
\end{equation}
with $W^0=1$ is included in $w_{1+\infty}$ (as opposed to $w_\infty$). As aforementioned, the $W^s$ correspond to higher-spin fields under $SL(2,\mathbb{R})$ \footnote{To compare with the generators of \cite{Guevara:2021abz} we take $W^{s}=w^{1+s/2}$. See Appendix I.}.
Their algebra is also $SL(2,\mathbb{R})$ covariant:
\begin{equation}
[W^m_{\alpha_{1}\cdots\alpha_{m}},W^n_{\beta_{1}\cdots\beta_{n}}]{=}i\tau mn\epsilon_{(\beta_{1}(\alpha_{1}}W^{m+n-2}_{\alpha_{2}{\cdots}\alpha_{m})\beta_{2}\cdots\beta_{n})}+\mathcal{O}(\tau^{2})\label{eq:Walg}
\end{equation}
This construction is closely
related to classical twistor space \cite{PENROSE197765,Penrose:1987uia}, locally $\mathbb{CP}^{1}\times\mathbb{C}^{2}$,
where $\mu_{+},\mu_{-}$ are simply coordinates on $\mathbb{C}^{2}$.
This motivates us to connect to momentum space through a Fourier transform
$\{\mu_{\alpha}\}\to\{\lambda_{\alpha}\}$. Thus we introduce the
momentum wavefunctions
\begin{align}
G^{\lambda} & :=e^{[\lambda\mu]}=\sum_{s=0}^{\infty}\frac{1}{s!}\lambda^{\alpha_{1}}\cdots\lambda^{\alpha_{s}}W^s_{\alpha_{1}\cdots\alpha_{s}}\,\,\,,\,\lambda_{\alpha}\in\mathbb{R}^{2}\,. \label{eq:RSFF}
\end{align}
Remarkably, in this basis the Weyl algebra (\ref{eq:Walg}) takes
the form
\begin{equation}
[G^{\lambda_{1}},G^{\lambda_{2}}]=i\tau[\lambda_{1}\lambda_{2}]G^{\lambda_{1}+\lambda_{2}}+\mathcal{O}(\tau^{2})\,.\label{eq:eag}
\end{equation}
This is precisely what appears in the numerator of (\ref{eq:mfsad})
and can be interpreted as structure constants of $SU(N)$ when $N\to\infty$
as noted in the seminal work of \cite{Hoppe:1988gk}. As a warm up for the next section let us see how this proceeds.
First note that (\ref{eq:gsa}) admits only infinite dimensional representations:
Indeed the enveloping algebra (\ref{eq:Walg}) is essentially the
linear space of regular functions of $\mu_{+},\mu_{-}$ equipped with symplectic
form (\ref{eq:gsa}). To connect to $SU(N)$ or $U(N)$ we will instead obtain finite,
$N-$dimensional, representations. For this we ``exponentiate''
the algebra by introducing $g_{\pm}=e^{\mu_{\pm}}$ so that (\ref{eq:gsa}) 
becomes
\begin{equation}
g_{+}g_{-}=q\,g_{-}g_{+}\,\,\,,q:=e^{i\tau}\,,\label{eq:frs}
\end{equation}
where in the classical limit $q\to1$ we see that $g_{\pm}$ are c-numbers
in $\mathbb{R}^{2}$. Let now $N$ be an odd integer. If we set 
\begin{equation}
\tau=4\pi/N\,,
\end{equation}
it essentially follows from (\ref{eq:frs}) that $g_{\pm}^{N}=1$ \footnote{We assume that the center of the group is trivial.}.
Thus $g_{\pm}$ now live in a lattice $Z_{N}\times Z_{N}$ with modular
parameter $i\tau=4\pi i/N$. This is where $U(N)$ acts. More precisely,
there exist $N\times N$ matrices $g_{\pm}$ satisfying (\ref{eq:frs}), which is a Heisenberg group. The generators are then

\begin{equation}
\frac{\textbf{G}^{\lambda}}{4\pi i/N}:=e^{[\lambda\mu]}=q^{\lambda^{+}\lambda^{-}/2}g_{+}^{\lambda^{+}}g_{-}^{\lambda^{-}}\,\,\,,\,\,\lambda_{\alpha}\in Z_{N}\times Z_{N}\,.
\end{equation}
Thus in this language the $U(N)$ generators $\textbf{G}^{\lambda}$
carry an adjoint color index given by a 2-spinor $\lambda$! (To get $SU(N)$, which is associated to $w_\infty$ instead of $w_{1+\infty}$,  we simply discard $\mathbf{G}^{(0,0)}=W^0=1$). Moreover, the quantum group
relation (\ref{eq:frs}) implies the \textit{Berezin formula}
\begin{equation}
\textbf{G}^{\lambda_{1}}\textbf{G}^{\lambda_{2}}=\frac{q^{[12]/2}}{4\pi i/N}\textbf{G}^{\lambda_{1}+\lambda_{2}}\,.\label{eq:ggs}
\end{equation}
Indeed, by evaluating $[\textbf{G}^{\lambda_{1}},\textbf{G}^{\lambda_{2}}]$ from here one can precisely match the structure constants for $U(N)$ \cite{Hoppe:1988gk}. Since $\lambda^\alpha$ corresponds to momenta on a lattice, this realizes a two-dimensional color-kinematics duality in the sense of \cite{Cheung:2022vnd}. On the other hand, for generic, non-discrete, values of $\tau$, eq. \eqref{eq:ggs} still applies \footnote{It corresponds to the Moyal product familiar in the context of non-commutative geometry, see e.g. \cite{Seiberg:1999vs,Floratos:2005ij}. We thank K. Costello and R.
Monteiro for pointing this out.}, but the momenta is not quantized. In any case, as $N\to\infty,\tau\to0$ we obtain a  semiclassical limit, where the
momenta spinor $\lambda$ again lives in $\mathbb{R}^{2}$ and
\begin{equation}
\textbf{G}^{\lambda_{1}}\textbf{G}^{\lambda_{2}}=(N/4\pi i+[12]/2)\textbf{G}^{\lambda_{1}+\lambda_{2}}+\mathcal{O}(1/N)\,.
\end{equation}
The divergence $\sim N$ does not contribute to the commutator, which
becomes precisely (\ref{eq:eag}), using $\textbf{G}^{\lambda}\to i\tau G^{\lambda}$.
As a final remark, for discrete $N$ we have written the algebra in terms
of $N\times N$ matrices $g_{\pm}$. The operators $\mu_{\pm}$ are
defined as $\mu_{\pm}=\log g_{\pm}$ only in a formal sense and are
unbounded. From the 4d theory perspective this is expected since they
correspond to Goldstone modes as we now explain.

\section{4d kinematic algebra}

Having constructed 2-dimensional color space as parametrized by $\lambda_{\alpha}$,
it is easy to upgrade the construction to 4-dimensional kinematic
space. The motivation comes again from twistor theory where, at least
classically, the positions $\mu_{\alpha}\in\mathbb{C}^{2}$ are fibered
over $z=\eta_2/\eta_1 \in\mathbb{CP}^{1}$ \cite{PENROSE197765,Penrose:1987uia}. This motivates
us to construct the quantum operators from sections of this bundle,
i.e. operators $\mu_{\alpha}(z)$ satisfying
\begin{equation}
\mu_{\alpha}(z_{1})\mu_{\beta}(z_{2})\sim i\tau\frac{\epsilon_{\alpha\beta}}{z_{12}}\,,\label{eq:fmr}
\end{equation}
This is the affine extension of the 2d algebra (\ref{eq:gsa}) \footnote{It has also appeared as a worldsheet OPE in the related context of the twistor sigma model
\cite{Adamo:2021lrv,Adamo:2021zpw}.}. 
More importantly, recalling that $p=|\eta\rangle [\lambda|$ we can identify $z=\eta_2/\eta_1$ as living in the celestial sphere. As we explain in Appendix, this algebra generalizes the one previously introduced for
celestial Goldstone modes in the soft sector \cite{Himwich:2020rro}. Indeed here the operators $\mu_{\alpha}$
themselves can be interpreted as graviton Goldstone modes valid for
all energies. To see this we introduce the normal-ordered vertex operators
\begin{equation}
i\tau\mathbf{G}^{\lambda}(z)=:e^{[\lambda\mu(z)]}: \,,\label{eq:GOP}
\end{equation}
for which the analog of the Berezin formula (\ref{eq:ggs}) takes
the form 
\begin{align}
\mathbf{G}^{\lambda_{1}}(z_{1})\mathbf{G}^{\lambda_{2}}(z_{2}) & =(q^{[12]/z_{12}}/i\tau):e^{[\lambda_{1}\mu(z_{1})]+[\lambda_{2}\mu(z_{2})]}:\nonumber \\
 & \sim(q^{[12]/z_{12}}/i\tau)\mathbf{G}^{\lambda_{1}+\lambda_{2}}(z_{2})\nonumber \\
 & \sim\frac{[12]}{z_{12}}\mathbf{G}^{\lambda_{1}+\lambda_{2}}(z_{2})+\mathcal{O}(\tau)\,.\label{eq:ta1}
\end{align}
This mimics (\ref{eq:ggs}): In the second line we have taken the OPE limit $z_{12}\to0$ which discards the $\sim N$ terms, and in
the third line the classical limit $\tau=4\pi/N\to0$. An important comment is now in order. In the OPE
(\ref{eq:fmr}) both operators have conformal weight $1/2$ in $z$,
which requires $\lambda_{\alpha}$ to have weight $-1/2$. This is
familiar from the construction of gravitational Goldstone operators
\cite{Himwich:2020rro}. However, after the $\tau\to0$ limit is taken in (\ref{eq:ta1})
we can take $z$ and $\lambda$ to be independent, after all the latter
only enter through the structure constants of $U(\infty)$. We can
make this even more explicit by writing, at strict $\tau=0$,
\begin{equation}
\mathbf{G}^{\lambda_{1}}(z_{1})\mathbf{G}^{\lambda_{2}}(z_{2})\sim\frac{\int d^{2}\lambda_{3}f^{\lambda_{1},\lambda_{2}}_{\quad\,\,\,\,\lambda_{3}}}{z_{12}}\mathbf{G}^{\lambda_{3}}(z_{2})\,,\label{eq:fasa}
\end{equation}
where the kinematic structure constants
\begin{equation}
f^{\lambda_{1},\lambda_{2}}_{\quad\,\,\,\,\lambda_{3}}=f^{\lambda_{1},\lambda_{2},-\lambda_{3}}=[12]\delta^{2}(\lambda_{1}+\lambda_{2}-\lambda_{3})\,,
\end{equation}
are completely antisymmetric and correspond to the kinematic algebra \cite{Monteiro:2013rya}. Seen in this way, the OPE pertains Kac-Moody currents with
weight 1 under $SL(2,\mathbb{R})_{z}$, where the other $SL(2,\mathbb{R})_{\bar{z}}$
has been promoted to a color $w_{1+\infty}$ gauge group. From a CFT perspective, this reproduces the chiral OPE block of \cite{Guevara:2021abz} which 
resums all $SL(2,\mathbb{R})_{\bar{z}}$ descendants, see also Appendix I.

Furthermore, we can use standard arguments to show that associativity of the OPE \eqref{eq:fasa} immediately yields the \textit{kinematic Jacobi relation} of \cite{Monteiro:2013rya}
\begin{align}
    \int d^{2}\lambda (f^{\lambda_{1},\lambda_{2}}_{\quad\,\,\,\,\lambda}\,\,f^{\lambda,\lambda_3,\lambda_4} + f^{\lambda_{1},\lambda_{3}}_{\quad\,\,\,\,\lambda}\,\,f^{\lambda,\lambda_4,\lambda_2} +&f^{\lambda_{1},\lambda_{4}}_{\quad\,\,\,\,\lambda}\,\,f^{\lambda,\lambda_2,\lambda_3} )\nonumber \\
    &=0\,.\label{eq:jacb}
\end{align}
Note that this Lie algebra structure was completely invisible from the collinear factorization/S-Matrix perspective of \eqref{eq:gw}, in striking contrast to its OPE realization. Furthermore, given this Lie algebra it is natural to further incorporate 2d
primary fields of arbitrary helicity transforming in the adjoint representation
(\ref{eq:fasa}), i.e. 
\begin{equation}
\mathbf{G}^{\lambda_{1}}(z_{1})\mathbf{O}_{h}^{\lambda_{2}}(z_{2})\sim\frac{\int d^{2}\lambda_{3}f^{\lambda_{1},\lambda_{2}}_{\quad\,\,\,\,\lambda_{3}}}{z_{12}}\mathbf{O}_{h}^{\lambda_{3}}(z_{2}) \,,\label{eq:GMAt}
\end{equation}
which is precisely (\ref{eq:mfsad}) as promised. We elaborate on this result in the next section. Here let us point out that the algebraic analysis of refs. \cite{Himwich:2021dau,Mago:2021wje}, which studied the Jacobi identity in the celestial conformal basis is here synthesized in the Lie algebra relation \eqref{eq:jacb}, closer to the recent analysis of \cite{Ren:2022sws}. Such references also studied the different helicities case, which we discuss briefly in Appendix II.

\section{Tower of Soft Charges in Gravity}

Let us now provide example-based evidence that gravitational amplitudes can indeed be derived from the preceding Kac-Moody OPEs. We do so by computing correlation functions of soft modes. Indeed, as it is well known, a systematic study of gravitational scattering can be carried out through the IR expansion of the S-Matrix \cite{Weinberg:1965nx}, which is also easily realized as a conformal soft expansion in the celestial sphere \cite{Guevara:2019ypd,Arkani-Hamed:2020gyp,Himwich:2021dau}. Notably, as we further flesh out in the Appendix, this expansion of a graviton state amounts precisely to the decomposition given in eq. \eqref{eq:RSFF} and thus allow us to identify the tower of soft modes. In the operator language, it follows from \eqref{eq:GOP} that
\begin{equation}
   \mathbf{G}^{\lambda}(z)=\sum_n \frac{1}{n!}\lambda^{\alpha_1}\cdots \lambda^{\alpha_n}\mathbf{W}^n_{\alpha_1\cdots \alpha_n}(z)\,,\label{eq:GE12}
\end{equation}
where
\begin{equation}
    \mathbf{W}^n_{\alpha_1\cdots \alpha_n}(z) = \frac{1}{i\tau} :\mu_{\alpha_1}\cdots \mu_{\alpha_n}: (z) \,,\label{eq:GE13}
\end{equation}
is the 4d analog of \eqref{eq:2dws}. Very nicely, their affine algebra can be computed either directly from the Goldstone modes \eqref{eq:fmr} or by expanding both sides of \eqref{eq:ta1} in soft modes. At $\tau\to 0$ the result is 
\begin{align}
    \mathbf{W}^n_{\alpha_1{\cdots} \alpha_n}(z_1) \mathbf{W}^m_{\beta_1{\cdots }\beta_n}&(z_2)  \sim \nonumber \\
    &\frac{mn}{z_{12}}\epsilon_{(\beta_{1}(\alpha_{1}}\mathbf{W}^{m+n-2}_{\alpha_{2}{\cdots}\alpha_{m})\beta_{2}{\cdots}\beta_{n})}(z_2) \label{eq:gqsx}
\end{align}
This agrees with the algebra of \cite{Strominger:2021lvk,Ball:2021tmb,Adamo:2021lrv} as we show in Appendix I. Note here that the $\mathbf{W}^1\mathbf{W}^1 \sim \mathbf{W}^0$ OPE is precisely our starting eq. \eqref{eq:fmr}. Thus the operator $\mathbf{W}^0$ is required to generate the tower. We will recognize it as a central extension of the 4d Poincare algebra.

To understand correlation functions involving arbitrary helicity fields we consider \eqref{eq:GMAt}. It is convenient to rewrite it as an \textit{exponential soft theorem} \cite{Li:2018gnc,Hamada:2018vrw,Guevara:2019ypd,Bautista:2019tdr}, i.e.
\begin{equation}
:e^{[\lambda_1 \mu(z_1)]}:\mathbf{O}_{h}^{\lambda_{2}}(z_{2})\sim i\tau \frac{[12]}{z_{12}}e^{[\lambda_1\frac{\partial}{\partial\lambda_{2}}]}\mathbf{O}_{h}^{\lambda_{2}}(z_{2})\,, \label{eq:GMAt2}
\end{equation}
from where, expanding both sides in $\lambda_1^\alpha$, according to \eqref{eq:GE12}-\eqref{eq:GE13} immediately yields a tower of soft operator OPEs
\begin{align}
 \mathbf{W}^0(z_1) \mathbf{O}_{h}^{\lambda}(z_{2}) &= \textrm{regular}\\
    \mathbf{W}_{\alpha}^1(z_1) \mathbf{O}_{h}^{\lambda}(z_{2})& \sim  \frac{\lambda_{\alpha}}{z_{12}}\mathbf{O}_{h}^{\lambda}(z_{2}) \label{eq:momg}\\
  \mathbf{W}_{\alpha \beta}^2(z_1) \mathbf{O}_{h}^{\lambda}(z_{2})& \sim  2 \frac{\lambda_{(\alpha}\frac{\partial}{\partial \lambda^{\beta)}}}{z_{12}}\mathbf{O}_{h}^{\lambda}(z_{2}) \label{eq:agm}\\
   \mathbf{W}_{\alpha \beta \gamma }^3(z_1) \mathbf{O}_{h}^{\lambda}(z_{2})& \sim  3 \frac{\lambda_{(\alpha}\frac{\partial}{\partial \lambda^{\beta}}\frac{\partial}{\partial \lambda^{\gamma)}}}{z_{12}}\mathbf{O}_{h}^{\lambda}(z_{2}) \,, \label{eq:momg3}
\end{align}
etc... From here we confirm that $\mathbf{W}^0$ is trivial, consistent with the expectation that it has no scattering amplitudes. Moving on, we recognize $\mathbf{W}^1$ as a spacetime momentum: Introducing $\eta_a= (1\,\,z)$ the translation generator is \footnote{In Appendix we interpret $\mathbf{W}^1$ as a weight $3/2$ operator, in which case these are its global modes.}
\begin{equation}
    P_{a\alpha}=\frac{(i\tau)^{-1}}{2\pi i}\oint dz \,\eta_a (z) \mu_{\alpha} (z) \,\,\,\,\,\,a,\alpha=\pm \,.
\end{equation}
Contour-integrating in \eqref{eq:momg} this gives $[P_{a\alpha},\mathbf{O}^{\lambda}(z)]=\eta_a \lambda_{\alpha} \mathbf{O}^{\lambda}(z)$ as expected. Analogously, we recognize $\mathbf{W}^{2}=:\mu\mu:$ as the angular momentum generator $\lambda_{(\alpha}\frac{\partial}{\partial \lambda^{\beta)}}$ \cite{Witten:2003nn}. Similarly to the Sugawara construction, it can be reinterpreted as the stress-energy tensor in celestial CFT, see Appendix. The generator $\mathbf{W}^3$ closely resembles a conformal transformation \cite{Witten:2003nn}. 
\section{Soft theorems and recursion}
From the S-Matrix perspective, these generators are responsible for Weinberg soft theorems and their subleading extensions \cite{Weinberg:1965nx,Cachazo:2014fwa}, when inserted into generic correlation functions. Omitting $SL(2,\mathbb{R})$ indices for simplicity we have
\begin{align}
\langle\mathbf{W}^{r}(z)&\mathbf{O}^{\lambda_{1}}(z_{1})\cdots\mathbf{O}^{\lambda_{n}}(z_{n})\rangle \nonumber \\
&=\frac{1}{2\pi i}\oint\frac{dw}{w-z}\langle\mathbf{W}^{n}(w)\mathbf{O}^{\lambda_{1}}(z_{1})\cdots\mathbf{O}^{\lambda_{n}}(z_{n})\rangle\nonumber\\
&=\sum_{i=1}^{n}{\frac{\lambda_{i}\left(\frac{\partial}{\partial\lambda_{i}}\right)^{n{-}1}}{z-z_{i}}}\langle\mathbf{O}^{\lambda_{1}}(z_{1})\cdots\mathbf{O}^{\lambda_{n}}(z_{n})\rangle{+}\mathcal{P}^r(z) \label{eq:expsf}
\end{align}
where the first term controls the colinear factorizations that are present in the scattering amplitudes. The extra term $\mathcal{P}^r(z)$ denotes additional contributions: Multiparticle singularities and polynomial terms at $z\to \infty$. The former are likely associated to Virasoro descendants and we leave their study for future work. 

To gain further insight into this picture, let us write  \eqref{eq:expsf} as the soft theorems in \cite{He:2014bga,Guevara:2019ypd}.  Resorting to the celestial dictionary \cite{Pasterski:2016qvg}, the correlation functions in the LHS of \eqref{eq:expsf} correspond to the S-Matrix with a graviton insertion,
\begin{equation}
    \mathcal{A}_{n{+}1}(k^{+2},p_{1},\ldots,p_{n}) \leftrightarrow \langle \mathbf{G}^\lambda(z)\mathbf{O}^{\lambda_{1}}(z_{1})\cdots\mathbf{O}^{\lambda_{n}}(z_{n})\rangle \,.
\end{equation}
Using \eqref{eq:GE12} we can then translate the relation \eqref{eq:expsf} into 
\begin{align}
\mathcal{A}_{n{+}1}(k^{+2}&,p_{1},\ldots,p_{n})\nonumber\\
& =\sum_i \frac{(\epsilon^{+}\cdot p_{i})^{2}}{k\cdot p_{i}}e^{J_i}\mathcal{A}_{n}(p_{1},\ldots,p_{n}) +\ldots \,,\label{eq:agsd2}
\end{align}
where $J_i= [\lambda \frac{\partial}{\partial \lambda^i}]$ amounts to a Lorentz generator on particle $i$. We refer to \cite{Guevara:2019ypd} for further details, see also \cite{Hu:2022bpa}. The $\ldots$ denote the contribution from $\mathcal{P}^r$. Their contribution can be understood as follows. Let $\mathcal{A}_n$ and $A_n$ correspond to the dressed and stripped amplitude, respectively
\begin{equation}
    \mathcal{A}_n(p_i)= \delta^2\left(\sum_i \lambda_i\right) \delta^2\left(\sum_i z_i\lambda_i\right) A_n(p_i)\,.
\end{equation}
Then, it is easy to see from \eqref{eq:agsd2} that $\mathcal{A}_{n+1}$ develops the expected support on $\delta^2(\lambda+\sum_i \lambda_i)$, but not necessarily in $\delta^2(z\lambda+\sum_i z_i\lambda_i)$. The former is a consequence of our color $U(\infty)$ Lie algebra while the latter can be interpreted as a two-dimensional holomorphic contour, see discussion below. The presence of the polynomial $\mathcal{P}^r$ is precisely to enforce that contour. Moreover, in the case of pure graviton MHV amplitudes \eqref{eq:agsd2} defines a BCFW-type recursion relation $\mathcal{A}_n\to \mathcal{A}_{n+1}$, for we can show explicitly that $\mathcal{P}^r(z_0)$ is given by the operator
\begin{equation}
\sum_{i=1}^{n-1}\frac{[\lambda\lambda_{i}]}{z_{ni}}\sum_{j=0}^{r{-}4}\left(\frac{z_{0n}}{z_{in}}\right)^{j}\,\sum_{m=0}^{r{-}4{-}j}\binom{r-1}{m}(J_{i}{-}J_{n})^{r{-}1{-}m}J_{n}^{m}
\end{equation}
acting on the $n-$point amplitude $\mathcal{A}_n$. This illustrates that $\mathcal{P}^r(z_0)$ is a polynomial of degree $r-4$ in $z_0$. In particular $\mathcal{P}^r=0$ for $r\leq 3$, and thus $\mathbf{W}^1,\mathbf{W}^2,\mathbf{W}^3$ in \eqref{eq:expsf} generate the standard (sub)leading soft theorems \cite{Cachazo:2014fwa} as expected.

\section{Discussion and future directions}
In this work we have provided evidence that recent developments in \textit{celestial holography, color-kinematics duality}, and \textit{twistor theory} are intrinsically connected. A natural three-way bridge between these topics is given by the OPE language, which exposes the properties of a Kac-Moody algebra, such as associativity, completely concealed by the S-Matrix picture. On the other hand, a feature that those three topics have in common is extended symmetry algebras which point towards integrability of gravitational theories. From the CFT arena there are indeed known connections between $\mathcal{W}_{1+\infty}$ and other integrable structures such as the Yangian $Y(\mathfrak{gl}_1)$ \cite{Bernard:1990jw,Prochazka:2015deb,https://doi.org/10.48550/arxiv.1305.4100}. The situation is closely connected to the self-dual sector, an integrable instance where kinematic algebras emerge naturally \cite{DOLAN1982387,Dolan:1981fq,Monteiro:2011pc,Chacon:2020fmr}.

In a strict sense 
this is a level-0 theory which is therefore non unitary (this is expected for tree-level scattering). Moreover, the connection to self-dual gravity also suggests that there should not be a multiparticle S-Matrix due to integrability. Recent work has shown, however, that integrable theories can be reduced to non-trivial scattering in presence of defects or other auxiliary fields \cite{Costello:2022jpg,Costello:2022upu,Costello:2022wso}. This resonates with an old conjecture of Ward \cite{10.2307/37545,2003JMP....44.3147A} stating that self-dual systems yield rich dynamics under compactification.

The properties unveiled through celestial holography, such as IR modes and theirs soft theorems, also point towards a construction more general than the self-dual sector. As a step towards such, in Appendix II we introduce negative helicity graviton states, corresponding to a 2d scalar field $\mathbf{\Phi}^\lambda$. We find
\begin{equation}
    \langle\mathbf{G}^{\lambda_{1}}(z_{1})\mathbf{G}^{\lambda_{2}}(z_{2})\Phi^{\lambda_{3}}(z_{3})\rangle = \frac{[12]^6}{[13]^2[23]^2} \delta^4\left(\sum_i p_i\right) \,.
\end{equation}
In this formula four-momentum conservation emerges naturally from OPEs, contrasting with the usual CCFT picture \cite{Fan:2022vbz,Mizera:2022sln}. Here we interpret two delta functions as poles defining contours in $z$. In particular we find
\begin{equation} 
\oint\frac{dz_{1}dz_{2}}{2\pi i}\langle\mathbf{G}^{\lambda_{1}}(z_{1})\mathbf{G}^{\lambda_{2}}(z_{2})\Phi^{\lambda_{3}}(z_{3})\rangle 
  =f^{\lambda_{1},\lambda_{2},\lambda_{3}}\,.
\end{equation}
Moving on to the four-point function $\langle \mathbf{GGG}\Phi\rangle$, we find via the recursion that it does not vanish identically, but instead we have
\begin{equation}
    \oint\frac{dz_{1}dz_{2}dz_{3}}{2\pi}\langle\mathbf{G}^{\lambda_{1}}(z_{1})\mathbf{G}^{\lambda_{2}}(z_{2})\mathbf{G}^{\lambda_{3}}(z_{3})\Phi^{\lambda_{4}}(z_{4})\rangle
= \eqref{eq:jacb}
\end{equation}
Thus its residue vanishes identically only if the Jacobi identity holds! In any case, the correlation itself is non-zero but has support only when all points $z_i$ coincide. This is precisely the situation in twistor space, where $A_4(1^+ 2^+ 3^+ 4^-)$ is supported on a degree zero curve \cite{Witten:2003nn}! Further exploration of this fascinating direction is left for future work. 

Many questions regarding the nature of the proposed holographic theory remain open. The most pressing one pertains multiparticle singularities as well as the loop amplitudes, perhaps paralleling the developments of \cite{Costello:2022jpg,Costello:2022upu}. Relatedly, it is tempting to explore the finite $N$ regime as a fully quantum theory. In this case the naive OPE \eqref{eq:ta1} necessarily yields higher-spin operators as in string theory.

\subsection*{Acknowledgements}

We wish to thank the organizers of the conference \textit{Amplitudes 2022}, where this work was presented, for providing a stimulating environment for discussions and research. We also appreciate the kind hospitality and environment provided by the Institute for Advanced Studies, where this work was completed.

We thank Mina Himwich, Dan Kapec, Monica Pate, James Mangan, Sebastian Mizera, Ricardo Monteiro, Shu-Heng Shao, Andrew Strominger and Adam Tropper for enlightening conversations. We especially thank Kevin Costello and Ronald Bittleston for discussions on quantum groups and non-commutative geometry.
This work
was supported by DOE grant de-sc/000787 and the Black Hole Initiative at Harvard University,
which is funded by grants from the John Templeton Foundation and the Gordon and Betty Moore
Foundation. The author also receives support from the Harvard Society of Fellows.

\textbf{Note added}. While finalizing the writing stage of this paper we were informed of the parallel works \cite{Monteiro:2022lwm}-\cite{Bu:2022iak}, which have also discussed the color-kinematics duality in this context.

\bibliography{references}
\bibliographystyle{aipnum4-1}

\newpage
\onecolumngrid
\section*{Appendices}
\appendix
\section{
I. Relation to the conformal primary basis}\label{ap:I}

In celestial holography, the S-Matrix is cast into a boost eigenstate
basis that exhibits its 2d conformal symmetry \cite{Pasterski:2021rjz}. In the present
work we have in contrast attempted to formulate fundamental properties
of correlation functions strictly in momentum space. The latter idea
fits into a more general framework we shall elaborate further in \cite{Guevara:2023}.
In the meantime we will outline here one of its key points: namely
that both approaches are connected through the holomorphic soft expansion \cite{Cachazo:2014fwa}.

In scattering, incoming massless particles such as gravitons can be
represented by insertion of creation operators depending on an energy
scale $\omega$ and coordinates $z,\bar{z}$ \cite{Strominger:2017zoo}
\begin{equation}
a_{+2}^{\dagger}(\omega,z,\bar{z})\,\,\,\,,\,\,\omega>0\,,
\end{equation}
where $+2$ indicates the 4-dimensional helicity of a graviton. Here the
momenta $p=|\eta\rangle[\lambda|$ is parametrized as $|\eta\rangle=\sqrt{\omega}(1\,\,z)$
and $|\lambda]=\sqrt{\omega}(1\,\,\bar{z})$. They have an energetic
Laurent expansion starting by the IR singularity $\omega^{-1}$ in
tree level amplitudes, which we parametrize it as
\begin{equation}
a_{+2}^{\dagger}(\omega,z,\bar{z})=\sum_{k=0}^{\infty}\omega^{k-1}H_{1-k}(z,\bar{z})\,.\label{eq:elemn}
\end{equation}
With some abuse of notation, and following the lore of \cite{Strominger:2017zoo}, we interpret the operators $H_{1-k}(z,\bar{z})$ as inserting
states into 2d correlators rather than in the 4d S-Matrix. Equivalently, they can be recast as conformal primaries by introducing $G_{\Delta}^{+}=\int_{0}^{\infty}d\omega\omega^{\Delta-1}a_{+2}^{\dagger}(\omega,z,\bar{z})$
so that the soft modes in \eqref{eq:elemn} are extracted as $H_{k}=\textrm{Res}_{\Delta=k}G_{\Delta}^{+}$ \cite{Guevara:2019fsj}. These are
precisely the operators constructed in \cite{Guevara:2019fsj,Guevara:2021abz,Strominger:2021lvk}  at integer boost weight
$\Delta=h+\bar{h}=1-k$, and we refer to those references for more details. Now, since for a graviton $h-\bar{h}=2$ these states have

\begin{equation}
H_{1-k}(z,\bar{z}):\,\,(h,\bar{h})=\left(\frac{3-k}{2},-\frac{k+1}{2}\right)\,.\label{eq:weighr}
\end{equation}

We will consider split signature $(++--)$ where $z$ and $\bar{z}$
are independent and we do not distinguish between incoming/outgoing
operators (this amounts to an overall sign in the momenta). Accordingly,
the 4d Lorentz group is realized as $SL(2,\mathbb{R})_{z}\times SL(2,\mathbb{R})_{\bar{z}}$
acting independendtly on $z,\bar{z}$ with weights $(h,\bar{h})$.
Since $\bar{h}<0$ for $H_{k}$ (recall $k\leq1)$ it was proposed
in \cite{Guevara:2021abz} that the $H_{k}(z,\bar{z})$ admits a polynomial
expansion in $\bar{z}$. Labeling $k=4-2p$, where $p=\frac{3}{2},2,\frac{5}{2},\ldots$
the formulae given there can be written as 

\begin{equation}
H_{4-2p}(z,\bar{z})=\frac{1}{(2p-2)!}\sum_{n=1-p}^{p-1}\binom{2p-2}{p-n-1}w_{n}^{p}(z)\bar{z}^{p-1-n}\,,\label{eq:3fas}
\end{equation}
where $w_{n}^{p}(z)$ each have $SL(2,\mathbb{R})_{z}$ weight $3-p$,
as dictated by (\ref{eq:weighr}). On the other hand, it was found
in \cite{Strominger:2021lvk} that for fixed $p$ the $2p-1$ states can be nicely
assembled into a $SL(2,\mathbb{R})_{\bar{z}}$ multiplet of spin $p$.
Now, a very natural spin-$p$ representation of $SL(2,\mathbb{R})$
is given by totally symmetric rank $2p-2$ tensors (see e.g. \cite{Arkani-Hamed:2017jhn}),
so this suggests we can identify
\begin{equation}
w_{n}^{p}=\mathbf{W}_{\underbrace{+\cdots+}_{p-n-1}\underbrace{-\cdots-}_{p+n-1}}^{2p-2}\,, \label{eq:idnt}
\end{equation}
i.e. as the independent components of the $SL(2,\mathbb{R})$ tensor
$\mathbf{W}_{\alpha_{1}\cdots\alpha_{2p-2}}^{2p-2}$. Moreover, introducing
the spinor $|\hat{\lambda}]=(1\,\,\bar{z})$ we see that the combinatorial
factor of (\ref{eq:fasa}) gets reabsorbed and the sum is simply a
contraction:
\begin{equation}
H_{4-2p=1-k}(z,\bar{z})=\frac{1}{(2p-2)!}\mathbf{W}_{\alpha_{1}\cdots\alpha_{2p-2}}^{2p-2}(z)\hat{\lambda}^{\alpha_{1}}\cdots\hat{\lambda}^{\alpha_{2p-2}}\,.\label{eq:fssa}
\end{equation}
Since $\mathbf{W}^{2p-2}$ carry $SL(2,\mathbb{R})_{\bar{z}}$ indices, we
have made ``almost'' manifest the transformation properties of the
$w_{n}^{p}$ multiplet. The only caveat is that $|\hat{\lambda}]=(1\,\,\bar{z})$
is a inohomogeneous spinor and picks up a Jacobian factor when trasnforming
under $SL(2,\mathbb{R})_{\bar{z}}$. This is remedied easily by attaching
the energy factor, i.e. we can perform a little group transformation
such that
\begin{align}
|\eta\rangle & =\sqrt{\omega}(1\,\,z),|\lambda]=\sqrt{\omega}(1\,\,\bar{z})\nonumber \\
 & \longrightarrow|\eta\rangle=(1\,\,z),|\lambda]=\omega(1\,\,\bar{z})=\omega|\hat{\lambda}]\,. \label{eq:trasn}
\end{align}
Since the graviton $a_{+2}^{\dagger}(\omega,z,\bar{z})$ has helicity
$+2$ it picks up a factor $\omega^{2}$ under this transformation
and becomes, combining (\ref{eq:elemn}) and (\ref{eq:fssa}),
\begin{align}
\mathbf{G}^{\lambda=\omega\hat{\lambda}}(z) & =\omega^{2}a_{+2}^{\dagger}(\omega,z,\bar{z})\nonumber \\
 & =\sum_{k=0}^{\infty}\omega^{k+1}H_{1-k}(z,\bar{z})\nonumber \\
 & =\sum_{k=0}^{\infty}\omega^{k+1}\frac{1}{(k+1)!}\mathbf{W}_{\alpha_{1}\cdots\alpha_{k+1}}^{k+1}(z)\hat{\lambda}^{\alpha_{1}}\cdots\hat{\lambda}^{\alpha_{k+1}}\nonumber \\
 & =\sum_{k=0}^{\infty}\frac{1}{(k+1)!}\mathbf{W}_{\alpha_{1}\cdots\alpha_{k+1}}^{k+1}(z)\lambda^{\alpha_{1}}\cdots\lambda^{\alpha_{k+1}} \,,\label{eq:defGs}
\end{align}
which is nothing but the expansion induced from vertex operators \eqref{eq:GE12},
as long as we include $\mathbf{W}^{0}=1$ as a central term. A number of comments
are in order. First, we see from the third line that this is a soft
expansion in the (anti)holomorphic coordinate, namely in $\lambda=\omega\hat{\lambda}(\bar{z})$.
But this is precisely the holomorphic soft expansion that was used
in \cite{Cachazo:2014fwa} to derive the subleading soft theorems! Here it emerges
naturally to all orders in $\omega$ and we shall pursue further applications
in \cite{Guevara:2023}. Second, considering $SL(2,\mathbb{R})$ tensors, rather
than modes $w_{n}^{p}$, immediately bypassess the need to operate
with the light-transform \cite{Himwich:2021dau} of conformal primaries to obtain the
$w_{1+\infty}$ algebra. 

Why did the light-transform play a predominant role in previous discussions of $w_{1+\infty}$ \cite{Strominger:2021lvk,Himwich:2021dau}? This can be easily understood by resorting to the homogeneous formalism used in \cite{Guevara:2023} (see also \cite{Sharma:2021gcz}). In that reference the inverse light-transform in coordinates $|\lambda]=(1\,\,\bar{z}), |\eta]=(1\,\,\bar{z}')$ is given by
\begin{equation}
    H_{4-2p}(z,\eta)=\frac{1}{\Gamma(2p-1)}\oint \frac{[\lambda d\lambda]}{2\pi i} [\lambda \eta]^{2p-2} \mathbb{L}[H_{4-2p}](z,\lambda)\,\,\,,\,\,p>0\,,
\end{equation}
By comparing this with \eqref{eq:fssa} we immediately identify the Penrose transform
\begin{equation}
    \mathbf{W}_{\alpha_{1}\cdots\alpha_{2p-2}}^{2p-2} =  \oint \frac{[\lambda d\lambda] }{2\pi i}\lambda_{\alpha_1}\ldots \lambda_{\alpha_{2p-2}}  \mathbb{L}[H_{4-2p}](z,\lambda) \,.
\end{equation}
Further using the identification \eqref{eq:idnt}, using $[\lambda d\lambda]=d\bar{z}$, this is
\begin{equation}
    w^p_n (z) = \oint \frac{d\bar{z}}{2\pi i} \bar{z}^{p-n-1} \mathbb{L}[H_{2p-2}] (z,\bar{z})\,.
\end{equation}
Hence $w^p_n (z)$ are nothing but the modes of the light-transform $\mathbb{L}[H_{2p-2}]$! Indeed, our representation landed us straight
into the so-called wedge of the $w_{1+\infty}$ algebra derived in
\cite{Ball:2021tmb,Himwich:2021dau}
\[
w_{n}^{p}(z_{1})w_{m}^{q}(z_{2})\sim2\frac{m(p-1)-n(q-1)}{z_{12}}w_{m+n}^{p+q-2}(z_{2})\,\,\,,\,\,|m|\leq p-1 \,,
\]
which, by virtue of (\ref{eq:fssa}), can be shown to be equivalent
to the $SL(2,\mathbb{R})$ covariant form \eqref{eq:gqsx}. In particular we realize
that $L_{n}=\frac{1}{2}w_{n}^{2}(0)$ satisfy the $SL(2,\mathbb{R})$
algebra and act as
\begin{equation}
[L_{n},w_{m}^{q}]=(m-n(q-1))w_{m+n}^{q} \,,
\end{equation}
and so this defines a 2d energy-momentum tensor $\bar{T}(\bar{z})$
for the CFT, for after some work, it can be shown to be equivalent
to
\begin{equation}
\bar{T}(\bar{z})\mathbf{G}^{\lambda_{2}}(z_{2})\sim\frac{\bar{h}}{\bar{z}_{12}^{2}}\mathbf{G}^{\lambda_{2}}(z_{2})+\frac{1}{\bar{z}_{12}}\bar{\partial}_{2}\mathbf{G}^{\lambda_{2}}(z_{2}) \,, \label{eq:safa}
\end{equation}
where $\bar{T}(\bar{z}):=\frac{1}{2}\sum_{n}\frac{w_{n}^{2}(0)}{\bar{z}^{n+2}}$
and $\bar{h}=-\frac{1}{2}\omega\partial_{\omega}$ is the helicity
weight on $\lambda_{2}=\omega\hat{\lambda}$. As argued in \cite{Kapec:2016jld},
see also \cite{Pasterski:2022djr,Fan:2020xjj}, $\bar{T}$ is essentially the Shadow (or light) transform
of $H_{0}(z,\bar{z})=\frac{1}{2}\sum_{n}\binom{2}{1-n}\frac{w_{n}^{2}(z)}{\bar{z}^{n-1}}$,
for which (\ref{eq:safa}) is equivalent to

\[
H_{0}(z_{1,}\bar{z}_{1})\mathbf{G}^{\lambda_{2}}(z_{2})\sim\frac{\bar{z}_{12}^{2}\omega\partial_{\omega}+\bar{z}_{12}\bar{\partial}_{2}}{z_{12}}\mathbf{G}^{\lambda_{2}}(z_{2}) \,,
\]
which in turn is nothing but \eqref{eq:agm} written in the coordinates $\lambda_{2}=\omega\hat{\lambda}(\bar{z})$.

\subsubsection*{Relation to the PC system in the soft sector}

As a follow up of the preceding discussion it is worth comenting on the relation with the gravitational Goldstone operator $C(z,\bar z)$ introduced in \cite{Himwich:2020rro} for the leading soft mode. Denoting $P(z)=\bar{\partial}H_1(z,\bar{z})$ that reference studied the system
\begin{equation}
 P(z) C(w,\bar{w})\sim \frac {i}{z-w} \,, \label{eq:pcs}
\end{equation}
This then allowed to construct \textit{hard} states in CCFT as the dressings
\begin{equation}
    \mathcal{O}_h(\omega,z,\bar{z})= e^{\omega C(z,\bar{z)}}  \mathcal{\hat{O}}_h (\omega,z,\bar{z})
\end{equation}
where $\mathcal{\hat{O}}_h $ does not interact with soft gravitons. By direct comparison, we can identify the $PC$ algebra with our $\{\mu_+,\mu_-\}$ system \eqref{eq:fmr}. Indeed, from eqs. \eqref{eq:fssa}-\eqref{eq:GE13}, we see that in our construction 
\begin{equation}
    H_1(z,\bar{z})= \mu_+ (z) + \bar{z} \mu_- (z) \Longrightarrow P(z)= \mu_- (z)\,.
\end{equation}
On the other hand, we have also constructed hard gravitons (to all orders in energy) as \eqref{eq:GOP}, and thus we can identify
\begin{equation}
    :e^{[\mu \lambda]}: \leftrightarrow e^{\omega C(z,\bar{z})} \,,\label{eq:igssf}
\end{equation}
in the soft sector. Furthermore the leading soft-theorem derived there 
\begin{equation}
    P(z)  e^{\omega C(w,\bar{w})} \sim 
     \frac{i\omega}{z-w} e^{\omega C(z,\bar{z})} \,, \label{eq:dsxz}
\end{equation}
is nothing but \eqref{eq:momg} of the main text. We can see that identifying $C(z,\bar{z})=\mu_+ + \bar{z} \mu_- =  H_1$ is consistent with \eqref{eq:pcs}, \eqref{eq:igssf} and \eqref{eq:dsxz}, but also generalizes the Goldstone mode to all the orders in the soft expansion, i.e. not only to $H_1$ but for all $H_{1-k\,\,},k\geq 1$. Dressed states of the form $ O_h^\lambda(z) = e^{[\lambda \mu]}\hat{O}_h (z)$ satisfy the tower of soft theorems \eqref{eq:momg}-\eqref{eq:momg3}, but may not be the most general states to do so. The difference with the framework of \cite{Himwich:2020rro} relies mainly in the level terms $\langle CC\rangle$ which are fixed from the IR divergence. The identification $C=H_1$ done here precludes us from incorporating them and thus may be corrected at loop level. It would be fascinating to work out this connection.

\section{II. Negative helicity states and one-minus amplitudes}\label{sec:ap2}

Here we introduce negative helicity states that pair to $\mathbf{G}^{\lambda}(z)$
through two-point functions. Indeed, to compute non-trivial correlations
explicitly we need to specifiy the identity terms in the OPE. First,
we can read off the color 2-point structure in momentum space from
the $U(N)$ matrices in \eqref{eq:ggs}. We get
\begin{equation}
d^{\lambda_{1},\lambda_{2}}:=\frac{1}{N(i\tau)^{2}}\textrm{Tr}(\textbf{G}^{\lambda_{1}}\textbf{G}^{\lambda_{2}})=\delta^{2}(\lambda_{1}+\lambda_{2})\,,
\end{equation}
which is luckily diagonal as the usual CFT inner product. However,
this does not appear as a singularity in the $\mathbf{G}^{\lambda}(z)\mathbf{G}^{\lambda}(z)$
OPE \eqref{eq:ta1}, in turn leading to their interpretation as zero-level Kac-Moody
currents. This is consistent at tree-level, since a vanishing level
leads to vanishing of correlation functions involving only $\mathbf{G}^{\lambda}(z)$
operators, which should correspond to all-plus amplitudes, $\langle\mathbf{G}^{\lambda}\cdots\mathbf{G}^{\lambda}\rangle=0$. 

To obtain amplitudes with insertion of negative helicity gravitons,
we can include a field $\Phi^{\lambda}$ with chiral OPE
\begin{equation}
\mathbf{G}^{\lambda_{1}}\Phi^{\lambda_{2}}\sim\delta(z_{12})d^{\lambda_{1},\lambda_{2}}+\frac{\int d\lambda_{3}f_{\qquad\lambda_{3}}^{\lambda_{1},\lambda_{2}}}{z_{12}}\Phi^{\lambda_{3}}\,,\label{eq:gqs}
\end{equation}
where we recall the second term follows from universal colinear singularities
and thus holds for generic helicities. The first term is a contact
interaction which requires the theory to be defined in Lorentzian
2d signature (i.e. split 4d signature). Since in this picture $\mathbf{G}^{\lambda}(z)$
has weight one, it also entails that $\Phi^{\lambda}(z)$ must transform
as a scalar under $SL(2,\mathbb{R})_{z}$. From the Kac-Moody perspective
the emergence of such a colored scalar is natural: One can consider
a level term between the $(1,0)$ current $J^{a}$ and the Shadow
of the $(0,1)$ current $\bar{J}^{b}$, namely $J^{a}S[\bar{J}^{b}]\sim1/z_{12}^{2}d^{ab}$
\cite{Nande:2017dba}. This is equivalent to $J^{a}\bar{J}^{b}=\delta^{2}(z_{12})d^{ab}$,
or to $J^{a}\Phi^{b}=\delta(z_{12})d^{ab}$ in Lorentzian signature,
if we identify $\Phi$ with the light-transform of $\bar{J}$. 

Let us further justify the contact term from the S-Matrix perspective.
Negative helicity gravitons are canonically conjugate to positive
ones. In the notation of the previous Appendix, the disconnected S-Matrix
is given by the two-point function
\begin{equation}
\langle a_{-2}(\omega_{2},z_{2},\bar{z}_{2})a_{+2}^{\dagger}(\omega_{1},z_{1},\bar{z}_{1})\rangle=\frac{\delta(\omega_{1}+\omega_{2})}{\omega_{1}}\delta(z_{1}-z_{2})\delta(\bar{z}_{1}-\bar{z}_{2})\,,\label{eq:aat}
\end{equation}
up to an irrelevant normalization. Recall now we introduce homogeneous
coordinates via $\lambda=\omega\hat{\lambda}(\bar{z})$. Since $a_{-2}$
has 4d helicity $-2$ under the transformation \eqref{eq:trasn}, the appropriate
operator becomes $\Phi^{\lambda}(z)=\omega^{-2}a_{-2}(\omega,z,\bar{z})$,
c.f. \eqref{eq:defGs}. All together, the $\omega^{2}$ Jacobian cancels out
in (\ref{eq:aat}) and so we obtain

\begin{align}
\langle\mathbf{G}^{\lambda_{1}}(z_{1})\Phi^{\lambda_{2}}(z_{2})\rangle & =\delta(\omega_{1}+\omega_{2})\delta(z_{1}-z_{2})\delta(\omega_{1}\bar{z}_{1}+\omega_{2}\bar{z}_{2})\nonumber \\
 & =\delta(z_{12})\delta^{2}(\lambda_{1}+\lambda_{2})=\delta(z_{12})d^{\lambda_{1},\lambda_{2}}\,,
\end{align}
precisely as anticipated. This thus manifests the color structure
at the level of two-point functions. The next task is to compute the
anti-MHV amplitude $\langle \mathbf{GG}\Phi\rangle$, which is slightly
different from the usual computation due to the contact terms. Consider
the limits $z_{12}\to0$ and $z_{23}\to0$. Using the $\mathbf{GG}$
OPE \eqref{eq:ta1} together with (\ref{eq:gqs}) we find, respectively,

\begin{align}
\langle\mathbf{G}^{\lambda_{1}}(z_{1})\mathbf{G}^{\lambda_{2}}(z_{2})\Phi^{\lambda_{3}}(z_{3})\rangle & \sim\frac{[12]}{z_{12}}\delta(z_{23})\delta^{2}(\lambda_{1}+\lambda_{2}+\lambda_{3})\nonumber \\
 & \sim\frac{[23]}{z_{23}}\delta(z_{12})\delta^{2}(\lambda_{1}+\lambda_{2}+\lambda_{3})\,.
\end{align}
The two limits are compatible if we identify $\frac{1}{z}\leftrightarrow2\pi i\delta(z)$.
This would render the function $\langle\mathbf{G}^{\lambda_{1}}\mathbf{G}^{\lambda_{2}}\Phi^{\lambda_{3}}\rangle$
to be genuinely holomorphic, while entailing that it is only defined
on the support of a certain contour integral. Indeed, in the absence
of other kinematic singularities, the two-dimensional delta function
$\delta(z_{12})\delta(z_{23})$ is the natural contour prescription
for $z_{1},z_{2}$. After integration we get
\begin{align}
\oint\frac{dz_{1}dz_{2}}{2\pi i}\langle\mathbf{G}^{\lambda_{1}}(z_{1})\mathbf{G}^{\lambda_{2}}(z_{2})\Phi^{\lambda_{3}}(z_{3})\rangle & =[12]\delta^{2}(\lambda_{1}+\lambda_{2}+\lambda_{3})\nonumber \\
 & =f^{\lambda_{1},\lambda_{2},\lambda_{3}}\,.\label{eq:ddre}
\end{align}
(The antisymmetry of $f^{\lambda_{1},\lambda_{2},\lambda_{3}}$ is
reflected in the contour prescription, e.g. $\oint\frac{dz_{1}dz_{3}}{2\pi i}\langle\cdots\rangle=-\oint\frac{dz_{1}dz_{2}}{2\pi i}\langle\cdots\rangle$,
etc.). For real-valued $z_{i}$, corresponding to the 2d Lorentzian
slice, we have
\begin{align}
\frac{1}{2\pi i}\langle\mathbf{G}^{\lambda_{1}}(z_{1})\mathbf{G}^{\lambda_{2}}(z_{2})\Phi^{\lambda_{3}}(z_{3})\rangle & =f^{\lambda_{1},\lambda_{2},\lambda_{3}}\delta(z_{23})\delta(z_{12})\nonumber \\
 & =[12]\delta(z_{23})\delta(z_{12})\delta^{2}(\lambda_{1}+\lambda_{2}+\lambda_{3})\nonumber \\
 & =\left(\frac{[12]^{3}}{[13][23]}\right)^{2}\delta^{2}(\lambda_{1}+\lambda_{2}+\lambda_{3})\delta^{2}(z_{1}\lambda_{1}+z_{2}\lambda_{2}+z_{3}\lambda_{3})\nonumber \\
 & =A_{3}(p_{i})\delta^{4}(\sum_{i=1}^{3}p_{i})\,,\label{eq:p3gs}
\end{align}
which is the desired 3-point dressed S-Matrix. Note that momentum
conservation has emerged from both the color and the holomorphic structure. 

We now aim to study $\langle\mathbf{G}^{\lambda_{1}}\mathbf{G}^{\lambda_{2}}\mathbf{G}^{\lambda_{3}}\Phi^{\lambda_{4}}\rangle$.
Again, we consider first different OPE limits. Starting with $z_{12}\to0$
we obtain
\begin{equation}
\langle\mathbf{G}^{\lambda_{1}}(z_{1})\mathbf{G}^{\lambda_{2}}(z_{2})\mathbf{G}^{\lambda_{3}}(z_{3})\Phi^{\lambda_{4}}(z_{4})\rangle\to\frac{[12]}{z_{12}}\langle\mathbf{G}^{\lambda_{1}+\lambda_{2}}(z_{2})\mathbf{G}^{\lambda_{3}}(z_{3})\Phi^{\lambda_{4}}(z_{4})\rangle\,.\label{eq:dasvc}
\end{equation}
Consider first the Lorentzian case $z_{i}\in\mathbb{R}$. From (\ref{eq:p3gs})
the three-point function has support only on $z_{2}=z_{3}=z_{4}$.
Thus the OPE limit $z_{1}\to z_{2}$ implies that all four punctures
coincide, which is a very singular configuration from the point of
view of 4d kinematics. For generic kinematics (subjected to momentum
conservation) this singularity vanishes. Repeating the argument for
$z_{13},z_{14}\to0$ we see that the full correlation function has
no singularities for generic kinematics and hence must vanish (a purely
contact term is not allowed by conformal symmetry). In fact, this
statement is already made in the original twistor construction of
SYM amplitudes \cite{Witten:2003nn}! In that case the one-minus amplitude $A_{4}(1^{+}2^{+}3^{+}4^{-})$
only has support if the corresponding real twistor variables collapse
to a single point, and hence vanishes generically.

On the other hand, we have proposed here to interpret the holomorphic
momentum conservation as a two-dimensional contour prescription. Unlike
the 3-point case however, the function $\langle\mathbf{G}\mathbf{G}\mathbf{G}\Phi\rangle$
carries the additional singularities of the type (\ref{eq:dasvc})
and so it is natural to integrate it against a three-dimensional contour.
In fact, since the overal $SL(2,\mathbb{R})_{z}$ weight is 3, the
natural contour is
\begin{align}
\oint\frac{dz_{1}dz_{2}dz_{3}}{2\pi}\langle\mathbf{G}^{\lambda_{1}}(z_{1})\mathbf{G}^{\lambda_{2}}(z_{2})\mathbf{G}^{\lambda_{3}}(z_{3})\Phi^{\lambda_{4}}(z_{4})\rangle\label{eq:fsdasx}\\
=f_{\quad\,\,\,\,\lambda}^{\lambda_{1},\lambda_{2}}\,\,f^{\lambda,\lambda_{3},\lambda_{4}}+f_{\quad\,\,\,\,\lambda}^{\lambda_{1},\lambda_{3}}\,\,f^{\lambda,\lambda_{4},\lambda_{2}}+ & f_{\quad\,\,\,\,\lambda}^{\lambda_{1},\lambda_{4}}\,\,f^{\lambda,\lambda_{2},\lambda_{3}}\,,\nonumber \\
\nonumber 
\end{align}
which extends the 3-point statement (\ref{eq:ddre}). Here we have
employed the form of the singularities (\ref{eq:dasvc}) together
with the double residue (\ref{eq:ddre}). This combination vanishes
due to associativity \eqref{eq:jacb} and can be interpreted as a residue theorem
(see e.g. \cite{GarciaSepulveda:2019jxn} and references within for evaluation of multidimensional residues). This can be thought as the momentum space version of the argument given in \cite{Mizera:2019blq}, where the same residues where analyzed in the moduli space.
This is also precisely the condition found recently in momentum space
in \cite{Ren:2022sws}, which also extended the analysis to helicity-flipping terms.
For completeness we present here the OPE for general helicity
fields (the momentum-space version of the OPEs in \cite{Himwich:2021dau})
\begin{equation}
O_{h_{1}}^{\lambda_{1}}(z_{1})O_{h_{2}}^{\lambda_{2}}(z_{2})\sim\sum_{h}\kappa_{\quad h}^{h_{1}h_{2}}\frac{[12]^{h_{1}+h_{2}-h-1}}{z_{12}}O_{h}^{\lambda_{1}+\lambda_{2}}\,,
\end{equation}
where imposing the vanishing of the analog of (\ref{eq:fsdasx}),
or equivalently OPE associativity, leads to the constraints in $\kappa_{\quad h}^{h_{1}h_{2}}$
presented in \cite{Ren:2022sws,Mago:2021wje}. It would be interesting to find a kinematic Lie algebra realization in the cases where $h_2\neq h$, which a priori do not correspond to adjoint $U(\infty)$ representations.


\end{document}

%% file: author_list.tex

\author{Alfredo Guevara } \email[]{aguevaragonzalez@fas.harvard.edu}  \affiliation{Center for the Fundamental Laws of Nature, Society of Fellows, \& Black Hole Initiative,
Harvard University, Cambridge, MA 02138, USA}
%
%
 \noaffiliation
\vskip 0.25cm